\documentclass[aps,prb,twocolumn,showpacs,showkeys,footinbib,superscriptaddress]{revtex4-1}
\usepackage{amsmath}
\usepackage{amssymb}
\usepackage{graphicx}
\usepackage{bm}
\usepackage{color}
\usepackage{relsize}
\usepackage{braket}

\usepackage{hyperref}
\usepackage[all]{hypcap}

\renewcommand{\i}{\ensuremath{\mathrm{i}}}
\newcommand{\e}{\ensuremath{\mathrm{e}}}

\begin{document}
\title{Excitonic Stark effect in MoS$_2$ monolayers}
\author{Benedikt Scharf}
\email[]{benedikt.scharf@ur.de}
\affiliation{Department of Physics, University at Buffalo, State University of New York, Buffalo, NY 14260, USA}
\affiliation{Skolkovo Institute of Science and Technology, 3 Nobel St., Moscow, Russia 143026}
\author{Tobias Frank}
\affiliation{Institute for Theoretical Physics, University of Regensburg, 93040 Regensburg, Germany}
\author{Martin Gmitra}
\affiliation{Institute for Theoretical Physics, University of Regensburg, 93040 Regensburg, Germany}
\author{Jaroslav Fabian}
\affiliation{Institute for Theoretical Physics, University of Regensburg, 93040 Regensburg, Germany}
\author{Igor \v{Z}uti\'c}
\email[]{zigor@buffalo.edu}
\affiliation{Department of Physics, University at Buffalo, State University of New York, Buffalo, NY 14260, USA}
\author{Vasili Perebeinos}
\email[]{v.perebeinos@skoltech.ru}
\affiliation{Skolkovo Institute of Science and Technology, 3 Nobel St., Moscow, Russia 143026}

\date{\today}

\begin{abstract}
We theoretically investigate excitons in MoS$_2$ monolayers in an applied in-plane electric field. Tight-binding and Bethe-Salpeter equation calculations predict a quadratic Stark shift, of the order of a few meV for fields of 10 V/$\mu$m, in the linear absorption spectra. The spectral weight of the main exciton peaks decreases by a few percent with an increasing electric field due to the exciton field ionization into free carriers as reflected in the exciton wave functions. Subpicosecond exciton decay lifetimes at fields of a few tens of V/$\mu$m could be utilized in solar energy harvesting and photodetection. We find simple scaling relations of the exciton binding, radius, and oscillator strength with the dielectric environment and an electric field, which provides a path to engineering the MoS$_2$ electro-optical response.
\end{abstract}


\maketitle

\section{Introduction}
In the past decade, atomically thin two-dimensional (2D) layers have emerged as a class of very versatile materials with highly tunable electronic and optical properties. Among these materials, a finite direct band gap makes monolayers (MLs) of MoS$_2$ and other transition metal dichalcogenides\cite{Bromley1972:JPC,Mak2010:PRL,Wang2012:NN,Kormanyos2015:2DM} (TMDs) attractive candidates for possible applications in nanoscale electronics, optoelectronics, and energy harvesting.\cite{Lembke2012:ACSNano,Bao2013:APL,LopezSanchez2013:NN,Britnell2013:S,Pospischil2014:NN,Yin2014:S,Wang2015:NT,Cui2015:NN,Rathi2015:NL,Dumcenco2015:ACSNano}

Due to inversion symmetry breaking, combined with strong spin-orbit coupling (SOC), these materials show several peculiar properties such as valley-dependent optical selection rules that allow for an efficient control of the spin- and valley-degrees of freedom by optical helicity,\cite{Xiao2012:PRL,Cao2012:NC,Zeng2012:NN,Mak2012:NN} the valley Hall\cite{Mak2014:S} and valley Zeeman\cite{Srivastava2015:NP,Stier2016:NC} effects, as well as strong magneto-\cite{Scrace2015:NN} and photoluminescence\cite{Mak2010:PRL,Splendiani2010:NL} with a quantum yield that can exceed $95\%$.\cite{Amani2015:S} In the context of spintronics,\cite{Zutic2004:RMP,*Fabian2007:APS} based on the large difference between the spin relaxation times of electrons and holes in TMDs,\cite{Song2013:PRL,Yang2015:NP} MoS$_2$ has been predicted as a desirable active region for spin-lasers,\cite{Lee2014:APL} while hybrid structures of graphene on TMDs have been proposed as a platform for optospintronics.\cite{Gmitra2015:PRB}

One of the most intriguing aspects of ML TMDs is that the interplay of their 2D character and Coulomb interactions leads to pronounced many-body effects that also dominate their optical properties. Strong excitonic effects with binding energies of several hundreds of meV, orders of magnitude larger than in conventional 3D semiconductors, are predicted in TMDs.\cite{Cheiwchanchamnangij2012:PRB,Ramasubramaniam2012:PRB,Shi2013:PRB,Berkelbach2013:PRB,Qiu2013:PRL,Komsa2013:PRB,MolinaSanchez2013:PRB,Steinhoff2014:NL,Zhang2014:PRB,Wu2015:PRB,Stroucken2015:JPCM,Dery2015:PRB,Wang2016:PRB} Due to the peculiar 2D screening and band structure these excitons are, moreover, expected to deviate from a simple hydrogen model.\cite{Qiu2013:PRL,Srivastava2015:PRL,Zhou2015:PRL} These predictions, large binding energies and a nonhydrogenic Rydberg series, have recently been confirmed experimentally.\cite{Chernikov2014:PRL,Chernikov2015:PRL,He2014:PRL,Wang2015:PRL,Ugeda2014:NM,Ye2014:N,Zhu2015:NSR,Hanbicki2015:SSC,Poellmann2015:NM} Likewise, the binding energies of trions (charged excitons) are also much larger in TMDs than in conventional semiconductors, of the order of several tens of meV.\cite{Mak2013:NM,Ross2013:NC,Berkelbach2013:PRB,Zhang2014:PRB,Ganchev2015:PRL} Indirect excitons in TMD-based van der Waals heterostructures similarly possess large binding energies and can be controlled by an electrostatic gate voltage.\cite{Fogler2014:NC,Calman2016:APL}

\begin{figure}[t]
\centering
\includegraphics*[width=8.5cm]{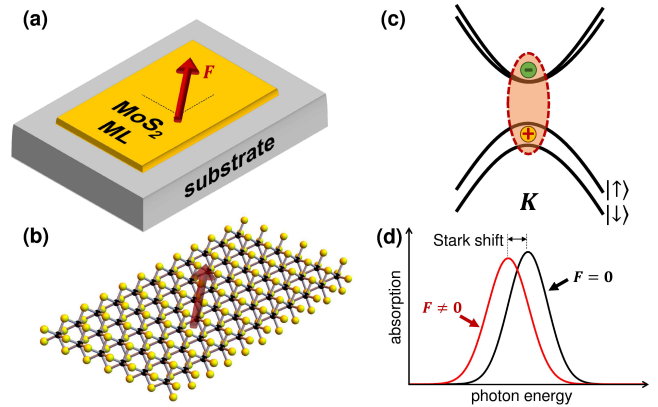}
\caption{(Color online) (a) Schematic setup of the system investigated, (b) MoS$_2$ monolayer (ML), (c) formation of the lowest-energy exciton, the so-called $A$ exciton, at the $K$ point in a MoS$_2$ ML, and (d) manifestation of the Stark effect on an exciton absorption peak if an electric field $\bm{F}$ is applied.}\label{fig:Scheme}
\end{figure}

In the present work, we study ML MoS$_2$ as the prototype material for ML TMDs and how excitons and the optical absorption in this system are affected by a constant in-plane electric field [Figs.~\ref{fig:Scheme}~(a,~b)]. Almost 60 years ago, Franz and Keldysh predicted a modulation of the band-edge absorption due to the electric field in bulk semiconductors.\cite{Franz1958:ZN,*Keldysh1958:JETP} If the effect of the Coulomb interaction between electrons and holes and, in consequence, excitons are taken into account [Fig.~\ref{fig:Scheme}~(c)], an excitonic Stark effect arises [Fig.~\ref{fig:Scheme}~(d)]: Similar to the hydrogen atom, there is a quadratic energy shift of the non-degenerate lowest energy exciton in the applied electric field. Recently, much theoretical attention has been paid to this effect\cite{Pedersen2016:NJoP,Haastrup2016:PRB,Pedersen2016:PRB}, for example, by addressing the problem of a zero radius of convergence and applying complex-scaling techniques\cite{Mera2015:PRL} which also yield the imaginary part of the excitonic resonance.

However, this effect is not easily observed in bulk semiconductors due to the low exciton binding energies exhibited in these systems. In contrast, as a consequence of their large exciton binding energies, TMDs enable probing an excitonic Stark effect, similar to quantum well structures,\cite{Miller1984:PRL,*Kuo2005:N} carbon nanotubes,\cite{Perebeinos2007:NL} or black phosphorus.\cite{Chaves2015:PRB} In fact, the Stark effect due to an out-of-plane electric field has recently been observed in mono- and few-layer TMDs,\cite{Withers2015:NL,Klein2016:NL,Vella2016:Arxiv,Matsuki2016:JJAP} while an a.c. optical Stark effect has been demonstrated in WSe$_2$ and WS$_2$, important for quantum information applications with TMDs.\cite{Kim2014:S,Sie2015:NM}

We predict a quadratic Stark shift with the in-plane field, which depends sensitively on the dielectric environment and is of the order of several meV or even larger for fields of a few tens of V/$\mu$m. Moreover, we provide the corresponding scaling laws for experimentally observable quantities such as the Stark shift and the loss of oscillator strength, demonstrating a tunable electro-optical response in MoS$_2$. While the focus of this work is on the linear absorption and thus on bright $s$-like excitons, we also predict a nonlinear Stark shift for the dark excitons.

\section{Theoretical Model}
We use an ab-initio-based tight-binding Hamiltonian that allows us to reproduce the single-particle band structure obtained by density functional theory (DFT). The linearized augmented plane wave code \textsc{wien2k}\cite{Blaha2014} is used employing the exchange-correlation functional PBE\cite{Perdew1996:PRL} to compute the DFT band structure of MoS$_2$ with a lattice constant of 3.14 \AA.\footnote{The energy cutoff is set to an equivalent of 150 eV to ensure converged results. Self-consistent calculations are done at a $k$-sampling of $30\times30$.} Then, the Wannierization is carried out with a $12\times12$ uniform sampling of the Brillouin zone using the \textsc{wannier90} package.\cite{Marzari1997:PRB} Depending on whether SOC is taken into account, a linear combination of S-centered $p$- and Mo-centered $d$-orbitals is chosen to project onto the separated manifold of either five (no SOC) or ten (SOC) bands in the low-energy region. The spreads of the maximally localized Wannier functions are smaller than 5 \AA$^2$.

A given single-particle state $\ket{n\bm{k}}$ with the 2D wave vector $\bm{k}$ in band $n$ and energy $\epsilon_{n\bm{k}}$ can then be written as the Bloch sum
\begin{equation}\label{BlochSum}
\ket{n\bm{k}}=\frac{1}{\sqrt{N}}\sum\limits_{\nu,i}\e^{\i\bm{k}\cdot\bm{R}_i}a_{n\bm{k}\nu}\ket{\nu,\bm{R}_i},
\end{equation}
where $\ket{\nu,\bm{R}_i}$ is the Wannier orbital $\nu$ centered at the Bravais lattice point $\bm{R}_i$ and $N$ is the total number of primitive unit cells considered. The coefficients $a_{n\bm{k}\nu}$ are determined from the tight-binding Hamiltonian $\mathcal{H}_{\nu\nu'}(\bm{k})$ via $\sum_{\nu'}\mathcal{H}_{\nu\nu'}(\bm{k})a_{n\bm{k}\nu'}=\epsilon_{n\bm{k}}a_{n\bm{k}\nu}$. Throughout this manuscript, we will label conduction and valence band indices by $n=c$ and $n=v$.

In the absence of an electric field, we employ the procedure described in Ref.~\onlinecite{Rohlfing2000:PRB} to compute excitons $S$ with momentum $q_\mathrm{exc}=0$ and solve the Bethe-Salpeter equation (BSE),
\begin{equation}\label{BSE}
\Delta_{vc\bm{k}}A^S_{vc\bm{k}}+\sum\limits_{v'c'\bm{k}'}\mathcal{K}_{vc\bm{k},v'c'\bm{k}'}A^S_{v'c'\bm{k}'}=\Omega_SA^S_{vc\bm{k}}.
\end{equation}
Here, $\Omega_S$ denotes the energy of the exciton state $\ket{\Psi_S}=\sum_{vc\bm{k}}A^S_{vc\bm{k}}\hat{c}^\dagger_{c\bm{k}}\hat{c}_{v\bm{k}}\ket{\mathrm{GS}}$ with the coefficients $A^S_{vc\bm{k}}$, the creation (annihilation) operator of an electron with momentum $\bm{k}$ in a conduction band $c$ (valence band $v$) $\hat{c}^\dagger_{c\bm{k}}$ ($\hat{c}_{v\bm{k}}$), and the ground state $\ket{\mathrm{GS}}$ with fully occupied valence bands and unoccupied conduction bands. The BSE is governed by the energy difference $\Delta_{vc\bm{k}}=\epsilon_{c\bm{k}}-\epsilon_{v\bm{k}}$ between the non-interacting\footnote{Throughout this work, we use single-particle instead of quasiparticle states. While calculations preformed with $GW$ give a more accurate description of the electronic structure of MoS$_2$,\cite{Qiu2013:PRL} the free-particle electron-hole pair energy enters in the diagonal of the BSE and thus the $GW$ bandgap renormalization does not change our conclusions regarding the Stark effect. Higher order effects entering via effective mass renormalization in $GW$ will modify the binding energy and hence the Stark shift. However, experimental uncertainty in the effective dielectric screening gives a much broader range of variation for the Stark effect.} states $\ket{c\bm{k}}$ and $\ket{v\bm{k}}$ and the interaction kernel
\begin{equation}\label{Kernel}
\mathcal{K}_{vc\bm{k},v'c'\bm{k}'}=\mathcal{K}^\mathrm{d}_{vc\bm{k},v'c'\bm{k}'}+\mathcal{K}^\mathrm{x}_{vc\bm{k},v'c'\bm{k}'},
\end{equation}
which consists of the direct and exchange terms, $\mathcal{K}^\mathrm{d}_{vc\bm{k},v'c'\bm{k}'}$ and $\mathcal{K}^\mathrm{x}_{vc\bm{k},v'c'\bm{k}'}$.\footnote{If the non-interacting single-particle/quasiparticle states are spin-degenerate, the exciton states can be categorized as singlet and triplet states, whose BSE~(\ref{BSE}) is calculated only from the real-space single-particle/quasiparticle wave functions and contains the kernels $\mathcal{K}=\mathcal{K}^\mathrm{d}+2\mathcal{K}^\mathrm{x}$ and $\mathcal{K}=\mathcal{K}^\mathrm{d}$, respectively.}

We model the interaction in $\mathcal{K}^\mathrm{d}_{vc\bm{k},v'c'\bm{k}'}$ by the screened Coulomb interaction in a 2D insulator,\cite{Keldysh1979:JETP,*Cudazzo2011:PRB,Berkelbach2013:PRB}
\begin{equation}\label{RPAPotential}
W(|\bm{R}^{\nu\nu'}_{ij}|)=\frac{e^2}{8\varepsilon_0 r_0}\left[H_0\left(\frac{\varepsilon|\bm{R}^{\nu\nu'}_{ij}|}{r_0}\right)-Y_0\left(\frac{\varepsilon|\bm{R}^{\nu\nu'}_{ij}|}{r_0}\right)\right],
\end{equation}
where $H_0$ and $Y_0$ are the Struve function and the Bessel function of the second kind. Here, $\bm{R}^{\nu\nu'}_{ij}=\bm{R}_{ij}+\bm{\tau}_\nu-\bm{\tau}_{\nu'}$, where $\bm{R}_{ij}=\bm{R}_i-\bm{R}_j$ and $\bm{\tau}_\nu$ and $\bm{\tau}_{\nu'}$ denote the centers of the Wannier orbitals $\nu$ and $\nu'$ in the primitive unit cell as computed by \textsc{wannier90}. The length $r_0=2\pi\chi_\mathrm{2D}$ is related to the 2D polarizability $\chi_\mathrm{2D}$,\cite{Berkelbach2013:PRB} $e=|e|$ is the absolute value of the electron charge, and $\varepsilon_0$ and $\varepsilon$ are the vacuum permittivity and the background dielectric constant. The background dielectric constant is given by $\varepsilon=(\varepsilon_1+\varepsilon_2)/2$, where $\varepsilon_{1,2}$ denotes the dielectric constants of the materials above and below the MoS$_2$ layer. This potential has proven highly successful in capturing excitonic properties of TMDs.\cite{Berkelbach2013:PRB}

Assuming point-like Wannier orbitals, the direct and exchange terms,\footnote{The exchange contribution is affected by the dielectric environment, but not by the screening due to electron-electron interaction.\cite{Rohlfing2000:PRB} However, in our calculations, the exchange term does not contribute in any significant way.}
\begin{equation}\label{DirectTerm}
\begin{aligned}
\mathcal{K}^\mathrm{d}_{vc\bm{k},v'c'\bm{k}'}=-\sum\limits_{\nu\nu'}&a^*_{c\bm{k}\nu}a_{c'\bm{k}'\nu}a_{v\bm{k}\nu'}a^*_{v'\bm{k}'\nu'}\\
&\times\left[\frac{1}{N^2}\sum\limits_{i,j}\e^{-\i(\bm{k}-\bm{k}')\cdot\bm{R}_{ij}}\;W(|\bm{R}^{\nu\nu'}_{ij}|)\right]
\end{aligned}
\end{equation}
and
\begin{equation}\label{ExchangeTerm}
\begin{aligned}
\mathcal{K}^\mathrm{x}_{vc\bm{k},v'c'\bm{k}'}=\sum\limits_{\nu\nu'}&a^*_{c\bm{k}\nu}a_{v\bm{k}\nu}a_{c'\bm{k}'\nu'}a^*_{v'\bm{k}'\nu'}\\
&\times\left[\frac{1}{N^2}\sum\limits_{i,j}V(|\bm{R}^{\nu\nu'}_{ij}|)\right]
\end{aligned}
\end{equation}
are computed in real space with the screened interaction~(\ref{RPAPotential}) and the bare Coulomb interaction $V(|\bm{R}^{\nu\nu'}_{ij}|)=e^2/(4\pi\varepsilon_0\varepsilon|\bm{R}^{\nu\nu'}_{ij}|)$, respectively.\footnote{We omit terms with $\bm{R}^{\nu\nu'}_{ij}=\bm{0}$. We have also checked the effect of a strictly on-site Hubbard energy $U$ for a range from 0 to 100 eV, but found that a finite $U$ does not significantly modify the binding energy. For $\varepsilon=1$, $\chi_\mathrm{2D}=6.5$ \AA, and $F=0$, for example, we obtain binding energies $E_\mathrm{b}=509$ meV with $U=20$ eV and $E_\mathrm{b}=511$ meV with $U=100$ eV as compared to $E_\mathrm{b}=508$ meV obtained from our approach.} An electric field $F$ along the (Bravais) unit direction $\bm{e}_1$ is accounted for by including the potential
\begin{equation}\label{ElectricFieldBSE}
\begin{aligned}
U_F(\bm{R}^{\nu\nu'}_{ij})=&eF\bm{R}^{\nu\nu'}_{ij}\cdot\bm{e}_1\\
&\times\tanh\left\{k_0\left[\frac{1}{4}-\left(\frac{\bm{R}^{\nu\nu'}_{ij}\cdot\bm{e}_1}{L}\right)^2\right]\right\},
\end{aligned}
\end{equation}
and adding it to $W(|\bm{R}^{\nu\nu'}_{ij}|)$ in Eq.~(\ref{DirectTerm}).\cite{Perebeinos2007:NL} To avoid numerical instabilities, the potential~(\ref{ElectricFieldBSE}) contains a smoothing factor, where $L$ is the length of the super cell along the $\bm{e}_1$-direction. The parameter $k_0$ controls how fast the electrostatic potential decays at the edge of the super cell. Implementation of the electrostatic potential using Eq.~(\ref{ElectricFieldBSE}) is a mathematical convenience to produce a periodic saw-tooth-type potential, which gives a linear dependence at small $|\bm{R}^{\nu\nu'}_{ij}|$ compared to the super cell size $L$. The potential is zero at the super cell boundaries $\bm{R}^{\nu\nu'}_{ij}\cdot\bm{e}_1=\pm L/2$, which ensures its periodicity, with the sign of the potential depending on the sign of $\bm{R}^{\nu\nu'}_{ij}\cdot\bm{e}_1$. A general in-plane field $\bm{F}$ can be considered by decomposing the field as $\bm{F}=F_1\bm{e}_1+F_2\bm{e}_2$ and adding Eq.~(\ref{ElectricFieldBSE}) for each direction to $W(|\bm{R}^{\nu\nu'}_{ij}|)$ in Eq.~(\ref{DirectTerm}). Since in our calculations, we find that the Stark effect on the main exciton absorption peaks is not sensitive to the direction of the in-plane field, we restrict ourselves to a field along the $\bm{e}_1$-direction and Eq.~(\ref{ElectricFieldBSE}).

\begin{figure}[t]
\centering
\includegraphics*[width=8.5cm]{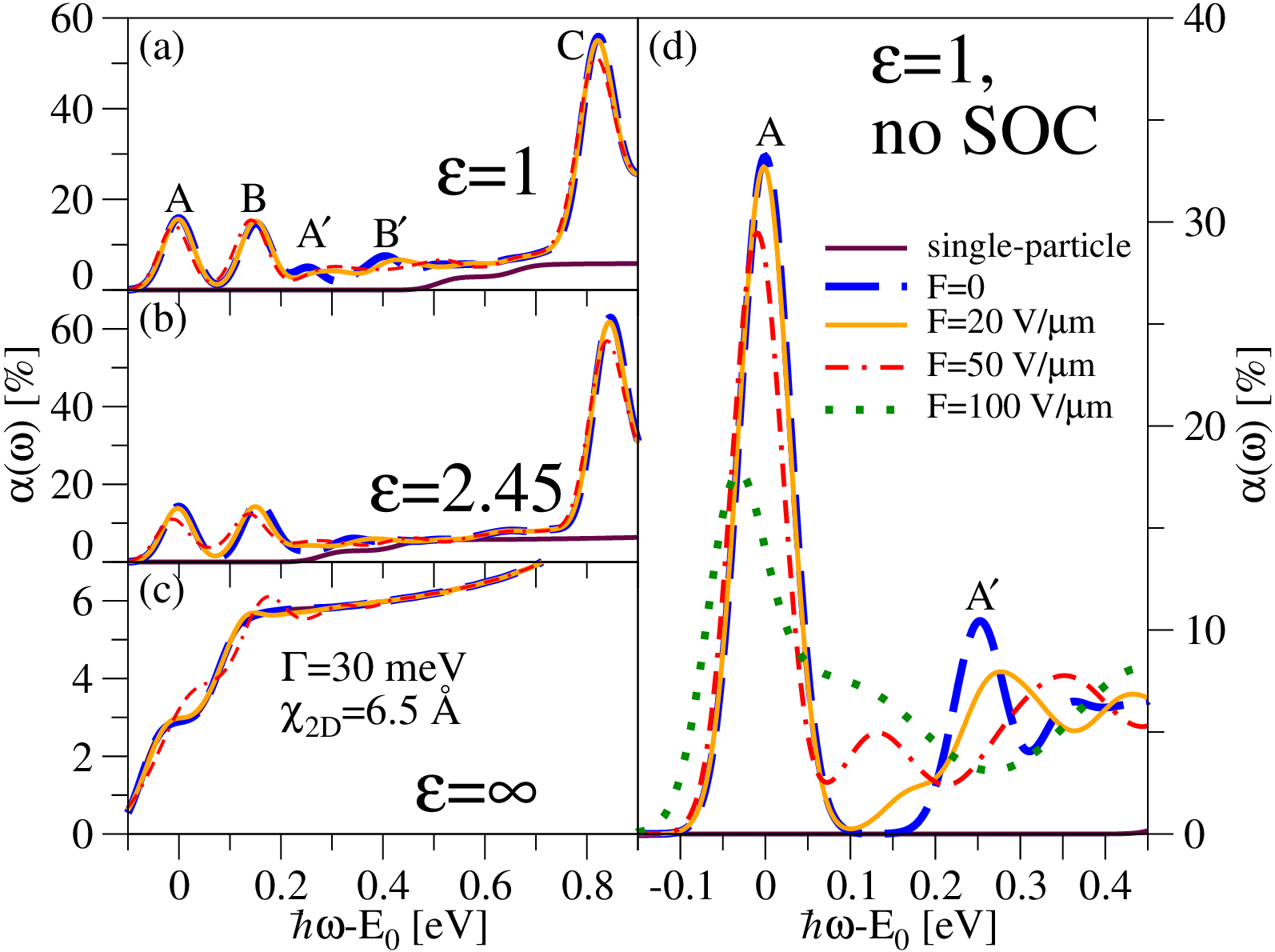}
\caption{(Color online) Calculated absorption spectra of monolayer MoS$_2$ for several different dielectric environments, (a) $\varepsilon=1$, (b) $\varepsilon=2.45$, and (c) $\varepsilon=\infty$, measured from the absorption onset $E_0$.\footnote{In experiments, $E_0\approx1.9$ eV. Here, we apply a rigid shift to the band structure obtained from our DFT-based tight-binding model to describe the quasiparticle band gap as obtained by GW.} (d) Same as in panel~(a), but without SOC. We use $\chi_\mathrm{2D}=6.5$ {\AA} and a Gaussian broadening of $\Gamma=30$ meV.}\label{fig:StarkAbsorption}
\end{figure}

The exciton states obtained from Eq.~(\ref{BSE}) can be used to compute the absorbance
\begin{equation}\label{absorption}
\alpha(\omega)=\frac{e^2\pi}{\varepsilon_0c\omega}\frac{1}{A}\sum\limits_S\left|\sum\limits_{vc\bm{k}}A^S_{vc\bm{k}}d_{vc}(\bm{k})\right|^2\delta(\hbar\omega-\Omega_S)
\end{equation}
of a 2D sheet with unit area $A$. Here, we have introduced the photon energy $\hbar\omega$, the single-particle/quasiparticle dipole-matrix element $d_{vc}(\bm{k})=\sum_{\nu,\nu'}a^*_{v\bm{k}\nu}a_{c\bm{k}\nu'}\partial_{k_x}\mathcal{H}_{\nu\nu'}(\bm{k})/\hbar$ obtained from the tight-binding model for the transition between states $\ket{c\bm{k}}$ and $\ket{v\bm{k}}$, and the velocity of light $c$. Since the gap between the conduction and valence bands is much larger than the spin-orbit splitting, we compute the excitons using the single-particle band structure without SOC and following Ref.~\onlinecite{Qiu2013:PRL} employ first-order perturbation theory to include SOC near the $K$ and $K'$ points.\footnote{Spin-orbit coupling is included by comparing the ab-initio band structures with and without SOC, adding their band- and $\bm{k}$-dependent energy differences as a perturbation to the BSE, and calculating the corresponding corrections in first-order perturbation theory.} Unless explicitly stated otherwise, our calculations are performed on a $144\times144$ $k$-grid/super cell with an upper energy cutoff of 2 eV above the band gap. This ensures that the exciton binding energies presented in this work are converged with a relative error of less than $10^{-5}$ when going from a $132\times132$ $k$-grid to a $144\times144$ $k$-grid. We have set the smoothing parameter $k_0=10$ throughout the manuscript and checked that the results are not affected for the fields presented in this work if larger values for $k_0$ are used ($k_0=20$, $k_0=30$, and $k_0=50$).

\begin{figure}[t]
\centering
\includegraphics*[width=8.5cm]{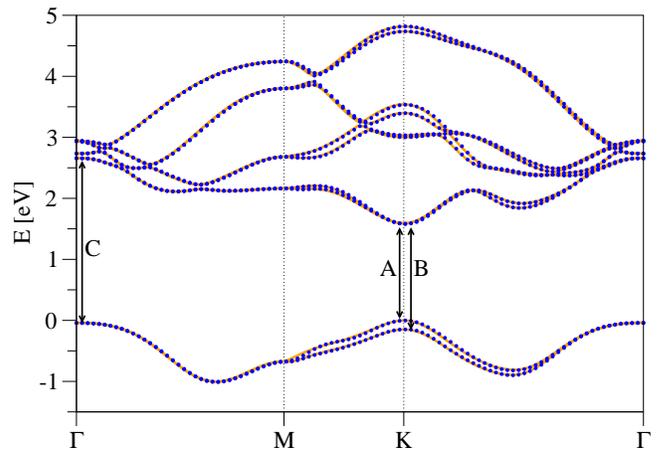}
\caption{(Color online) Band structure of monolayer MoS$_2$ as obtained from DFT (orange lines) and as calculated using an ab-initio-based tight-binding model (blue circles) with SOC. The optical transitions involving the $A$ and $B$ excitons at the $K$ ($K'$) point and the $C$ exciton near the $\Gamma$ point are also marked.}\label{fig:Dispersion}
\end{figure}

\section{Results}
To illustrate the effect of an electric field, Fig.~\ref{fig:StarkAbsorption} displays the absorption spectra of ML MoS$_2$ for several different dielectric environments. For clarity, we also show the behavior of the lowest exciton peaks for $\varepsilon=1$ in the absence of SOC in Fig.~\ref{fig:StarkAbsorption}~(d). At zero field and $\varepsilon=1$, one can clearly see the $A$ and $B$ exciton peaks originating from transitions into the spin-orbit-split valence bands at the $K$ and $K'$ points, their respective Rydberg states, $A'$ and $B'$, as well as the so-called $C$ exciton that arises due to transitions near the $\Gamma$ point [see Fig.~\ref{fig:StarkAbsorption}~(a)]. The origin of these excitons is also depicted in Fig.~\ref{fig:Dispersion}, which shows the band structure of ML MoS$_2$ obtained from our tight-binding description. Moreover, a comparison with the original DFT band structure illustrates an almost perfect agreement between the two band structures.

Due to the screened potential given by Eq.~(\ref{RPAPotential}) we obtain a nonhydrogenic Rydberg series for the $A$ and $B$ excitons, a fact well established experimentally\cite{Chernikov2014:PRL} and theoretically.\cite{Berkelbach2013:PRB,Qiu2013:PRL,Wu2015:PRB} Depending on the value of $\chi_\mathrm{2D}$ as given in the literature,\cite{Berkelbach2013:PRB} our model predicts binding energies $E_\mathrm{b}$ of around 500-600 meV for the $A$ and $B$ excitons at $\varepsilon=1$ (see also Table~\ref{tab:BindingEnergies}). As $\varepsilon$ is increased, the binding energies of the $A$, $B$, and $C$ excitons decrease, which can be seen in Fig.~\ref{fig:StarkAbsorption}~(b), where we have chosen $\varepsilon=(1+3.9)/2=2.45$ to model the dielectric environment of MoS$_2$ on a SiO$_2$ substrate.

If an electric field is applied, the binding energies of the $A$, $B$, and $C$ excitons increase by $\delta E_\mathrm{b}(F)=E_\mathrm{b}(F)-E_\mathrm{b}(F=0)$ due to the Stark effect. Their peaks, on the other hand, lose spectral weight, a part of which is transferred into the region between the $A$/$B$ exciton peaks and the onset of the continuum. At high fields an additional absorption peak arises in this region, while the continuum absorption exhibits Franz-Keldysh oscillations [see Figs.~\ref{fig:StarkAbsorption}~(a), (b) and~(d)]. The period of these oscillations is proportional to the electric field and their amplitude also grows with increasing field, which is corroborated in Fig.~\ref{fig:StarkAbsorption}~(c), where --- in the absence of any electron-hole interaction, $\varepsilon=\infty$, and consequently excitons --- the electric field leads only to Franz-Keldysh modulations of the absorption.\cite{Franz1958:ZN,*Keldysh1958:JETP}

\begin{widetext}
\begin{center}
\begin{table}
\begin{center}
\begin{tabular}{|c|c||c|c|c|c|}
\hline
$\chi_\mathrm{2D}$ [\AA] & $\varepsilon$ & $E_\mathrm{b}$ [meV] & $\kappa$ [(eV$\mu$m/V)$^2$] & $\tilde{\alpha}$ [eV$\mu$m$^2$/V$^2$] & $\lambda$ [($\mu$m/V)$^2$]\\
\hline\hline
5.0 & 1 & 603 & $2.43\times10^{-6}$ & $8.06\times10^{-6}$ & $2.99\times10^{-5}$\\
\hline
6.5 & 1 & 508 & $2.75\times10^{-6}$ & $1.08\times10^{-5}$ & $5.37\times10^{-5}$\\
\hline
6.5 & 2.45 & 279 & $2.28\times10^{-6}$ & $1.63\times10^{-5}$ & $1.20\times10^{-4}$\\
\hline
6.5 & 3.35 & 214 & $2.40\times10^{-6}$ & $2.24\times10^{-5}$ & $1.87\times10^{-4}$\\
\hline
6.5 & 5 & 143 & $2.53\times10^{-6}$ & $3.54\times10^{-5}$ & $4.21\times10^{-4}$\\
\hline
\end{tabular}
\end{center}
\caption{Binding energies $E_\mathrm{b}$, fitting parameters $\kappa$ with corresponding electric polarizabilities $\tilde{\alpha}$ in Eq.~(\ref{QuadraticStarkShift}), and $\lambda$ in Eq.~(\ref{QuadraticStarkShift_Oscillator}) for different dielectric environments $\varepsilon$ and $\chi_\mathrm{2D}$.}\label{tab:BindingEnergies}
\end{table}
\end{center}
\end{widetext}

\begin{figure}[t]
\centering
\includegraphics*[width=8.5cm]{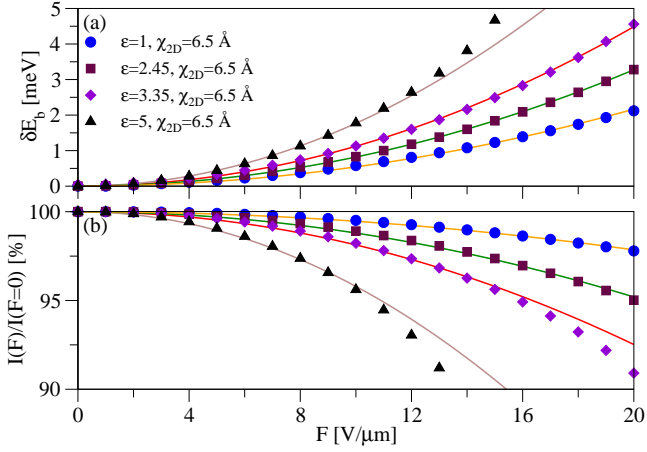}
\caption{(Color online) Field dependence of (a) the Stark shifts $\delta E_\mathrm{b}$ and (b) the relative oscillator strengths $I(F)/I(F=0)$ of the lowest excitons for different dielectric environments. Symbols denote computed data points, whereas the solid lines denote best fits given by Eqs.~(\ref{QuadraticStarkShift}) and~(\ref{QuadraticStarkShift_Oscillator}) and Table~\ref{tab:BindingEnergies}.}\label{fig:StarkShift}
\end{figure}

The Stark shift $\delta E_\mathrm{b}$ of the lowest excitonic states depends on $E_\mathrm{b}$ at zero field as shown in Fig.~\ref{fig:StarkShift}~(a). For fields up to $20$ V/$\mu$m, we find Stark shifts of the order of several meV, similar to the optical Stark shifts observed recently in WSe$_2$,\cite{Kim2014:S} while the shift is much larger for $F=50$ V/$\mu$m [see Figs.~\ref{fig:StarkAbsorption}~(a), (b) and~(d)], for example, $\delta E_\mathrm{b}\approx10$ meV for free standing MoS$_2$ ($\varepsilon=1$), $\delta E_\mathrm{b}\approx17$ meV for MoS$_2$ on SiO$_2$ ($\varepsilon=2.45$), and $\delta E_\mathrm{b}\approx18$ meV for MoS$_2$ on diamond [$\varepsilon=(1+5.7)/2=3.35$]. Figure~\ref{fig:StarkAbsorption}~(d) implies that for $\varepsilon\approx1$ the exciton peak can still be observed at $F=100$ V/$\mu$m with a Stark shift of $\delta E_\mathrm{b}\approx32$ meV. At low fields and high binding energies, $\delta E_\mathrm{b}$ can be fitted very well to a quadratic function
\begin{equation}\label{QuadraticStarkShift}
\delta E_\mathrm{b}=\tilde{\alpha}F^2/2=\kappa F^2/E_\mathrm{b},
\end{equation}
where the fitting parameter $\kappa$ is related to the electric polarizability $\tilde{\alpha}=2\kappa/E_\mathrm{b}$ and found to be around $\kappa\approx 2.5\times10^{-6}$ (eV$\mu$m/V)$^2$, nearly independent of $\varepsilon$ and $\chi_\mathrm{2D}$, with the actual values obtained for best fits given in Table~\ref{tab:BindingEnergies}. Our results found for $\tilde{\alpha}$ are of the same order as those recently computed in Ref.~\onlinecite{Pedersen2016:PRB}. Equation~(\ref{QuadraticStarkShift}) is motivated by the form of the second-order correction due to the Stark effect in the hydrogen atom and the small exciton radius. Since our calculations point to $\kappa$ being only weakly dependent on the dielectric background and the polarizabilities, Eq.~(\ref{QuadraticStarkShift}) implies that $E_\mathrm{b}$ in different dielectric setups can be obtained by fitting the quadratic field dependence of the Stark shift to a constant inversely proportional to $E_\mathrm{b}$. Figure~\ref{fig:StarkShift}~(a) also illustrates that for high fields and low binding energies, such as for $\varepsilon=5$, $\delta E_\mathrm{b}$ deviates from this quadratic behavior and higher order corrections in $F$ become more important.This breakdown of the quadratic behavior of $\delta E_\mathrm{b}$ at higher fields has also been observed in Ref.~\onlinecite{Haastrup2016:PRB} and roughly estimated in Ref.~\onlinecite{Pedersen2016:PRB} to happen at fields well below $F_\mathrm{b}=5.14\times10^{11} m_\mathrm{r}^2/\varepsilon^3$ V/m, where $m_\mathrm{r}$ is the reduced mass (in units of the electron mass) of the Wannier problem.\footnote{Here, we can estimate our reduced mass as $m_\mathrm{r}\approx0.25$. Since the estimate for $F_\mathrm{b}$ is taken from the hydrogen problem, whose binding energies scale differently than the binding energies in ML-TMDs, the actual breakdown of the quadratic approximation happens at much smaller fields, as can be seen in Fig.~\ref{fig:StarkShift} in our work for $\varepsilon=5$ or in Ref.~\onlinecite{Haastrup2016:PRB}.}

Moreover, we study the loss of the spectral weight of the exciton peaks with increasing $F$. Figure~\ref{fig:StarkShift}~(b) displays the oscillator strengths $I(F)$ [normalized to the oscillator strength at $F=0$, $I(F=0)$] of the lowest exciton peak, for which we find that its field dependence can be approximated by
\begin{equation}\label{QuadraticStarkShift_Oscillator}
\frac{I(F)}{I(F=0)}=1-\lambda F^2.
\end{equation}
Table~\ref{tab:BindingEnergies} gives values of the parameter $\lambda$, which varies widely for different combinations of $\chi_\mathrm{2D}$ and $\varepsilon$ and cannot be easily related to $E_\mathrm{b}$. Here, we have computed the oscillator strength as the sum $(1/N)\sum_S|\sum_{vc\bm{k}}A^S_{vc\bm{k}}d_{vc}(\bm{k})|^2$ of each single exciton peak $S$ contributing to the low energy peak, which is approximately proportional to the energy integral over the low energy peak.\footnote{Note that the absolute value of the oscillator strength $I$ defines a value of the radiative lifetime [R. Loudon, \textit{The Quantum Theory of Light, Third Edition (Oxford Science Publications)}, (Oxford Univ. Press, Oxford, 2000)], which we find to be $\tau_0\approx300$ fs for $\varepsilon=1$ and an experimental exciton energy of $1.9$ eV, consistent with the theoretical [P. San-Jose, V. Parente, F. Guinea, R. Rold\'an, and E. Prada, Phys. Rev. X \textbf{6}, 031046 (2016)] and experimental [T. Korn, S. Heydrich, M. Hirmer, J. Schmutzler, and C. Sch{\"u}ller, Appl. Phys. Lett. \textbf{99}, 102109 (2011)] reports.}

An electric field is expected to decrease the exciton lifetime leading to a broadening of the exciton absorption peak due to the exciton dissociation. These effects can be related to the loss of spectral weight displayed in Fig.~\ref{fig:StarkShift}~(b). Since this loss is quite small for typical values of $\varepsilon$, around $5\%$ for $\varepsilon=2.45$ and $F=20$ V/$\mu$m, Fig.~\ref{fig:StarkShift}~(b) implies that the $A$ and $B$ exciton peaks require very large fields beyond which they fully dissociate. This in turn suggests that one should be able to observe an excitonic Stark effect in MoS$_2$ MLs experimentally. The Rydberg states $A'$ and $B'$, on the other hand, dissociate already at smaller fields due to their lower binding energy as shown in Fig.~\ref{fig:StarkAbsorption}.

\begin{figure}[t]
\centering
\includegraphics*[width=8.5cm]{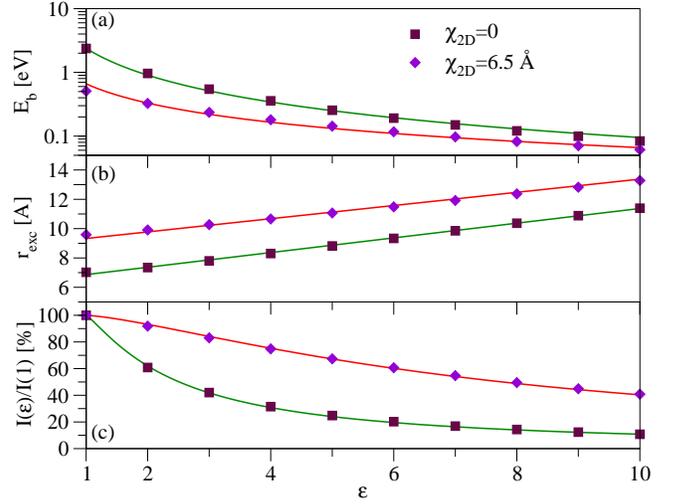}
\caption{(Color online) Dependence of (a) the binding energy $E_\mathrm{b}$, (b) the exciton radius $r_\mathrm{exc}$, and (c) the relative oscillator strengths $I(\varepsilon)/I(\varepsilon=1)$ of the lowest excitons at $F=0$. Symbols denote computed data points, whereas the solid lines denote best fits to a power law in panel~(a), to a linear dependence in panel~(b), and to Eq.~(\ref{DielectricConstant_Oscillator}) in panel~(c).}\label{fig:epsilon}
\end{figure}

As we have shown in Fig.~\ref{fig:StarkShift}, the Stark shifts strongly depend on the binding energy at zero field. Hence, we study the dependence of $E_\mathrm{b}$ of the lowest exciton on the dielectric environment in Fig.~\ref{fig:epsilon}, where we compare the results obtained for the screened potential with $\chi_\mathrm{2D}=6.5$ {\AA} with those of the bare Coulomb potential (with $\chi_\mathrm{2D}=0$). Figure~\ref{fig:epsilon}~(a) illustrates that the dependence of $E_\mathrm{b}$ on $\varepsilon$ can be reasonably well fitted to power laws, $E_\mathrm{b}\propto\varepsilon^{-1.4}$ for $\chi_\mathrm{2D}=0$ and $E_\mathrm{b}\propto\varepsilon^{-1}$ for $\chi_\mathrm{2D}=6.5$ {\AA}, which differ from the hydrogen model for the range of $\varepsilon$ in Fig.~\ref{fig:epsilon}. This is due to two reasons: (i) The discreteness of the Bravais lattice results in deviations from the continuum model of the hydrogen atom, which predicts the binding energy to scale as $E_\mathrm{b}\propto\varepsilon^{-2}$. (ii) For finite $\chi_\mathrm{2D}$, the potential deviates from the bare Coulomb potential at short distances $r\to0$, which is particularly relevant for small exciton radii $r_\mathrm{exc}$ and, hence, small $\varepsilon$. As we increase $\varepsilon$, $r_\mathrm{exc}$ becomes larger compared to the lattice constant and exponents closer to the hydrogen model are found, $E_\mathrm{b}\propto\varepsilon^{-1.7}$ and $E_\mathrm{b}\propto\varepsilon^{-1.5}$ for the bare and screened potentials, respectively, using a range of $\varepsilon$ from 10 to 20 (not shown).

The corresponding exciton radii $r_\mathrm{exc}$ of the lowest excitons are computed by fitting the exciton wave function (see below) to a 2D Gaussian with the standard deviation being used as an estimate for $r_\mathrm{exc}$ and are displayed in Fig.~\ref{fig:epsilon}~(b). One can see that $r_\mathrm{exc}$, although increasing with $\varepsilon$, does not change significantly for typical values of $\varepsilon$, by at most 10\% for $\varepsilon=5$. Figure~\ref{fig:epsilon}~(c) shows the oscillator strengths $I(\varepsilon)$ of the exciton peak normalized to its value at $\varepsilon=1$. With increasing $\varepsilon$, the spatial overlap of the electron and hole wave functions and, hence, $I(\varepsilon)$ are diminished as expected. In both cases, $\chi_\mathrm{2D}=0$ and $\chi_\mathrm{2D}=6.5$ {\AA}, its behavior [normalized to the oscillator strength $I(\varepsilon=1)$] scales with the exciton radius $r_\mathrm{exc}$ as
\begin{equation}\label{DielectricConstant_Oscillator}
\frac{I(\varepsilon)}{I(\varepsilon=1)}=\frac{\left[r_\mathrm{exc}(1)-R_0\right]^2}{\left[r_\mathrm{exc}(\varepsilon)-R_0\right]^2}
\end{equation}
with a length scale $R_0\approx|\bm{a}_{1/2}|=3.1$ {\AA} on the order of the lattice constant.

\begin{figure}[ht]
\centering
\includegraphics*[width=8.5cm]{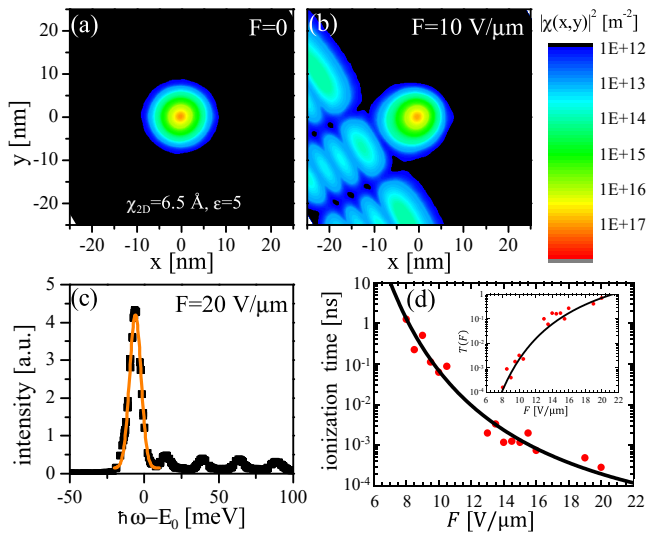}
\caption{(Color online) Density plots of the exciton wave function $|\chi(x,y)|^2$ at (a) $F=0$ and (b) $F=10$ V/$\mu$m in $x$-direction. (c) Optical transition intensities measured from the absorption onset $E_0$ at zero field (see also Fig.~\ref{fig:StarkAbsorption}) and Gaussian fit of the line shape at $F=20$ V/$\mu$m. (d) Ionization times as computed from the leakage of the wave function with the tunneling probabilities $T(F)$ shown in the inset. The solid black lines are fits according to Eq.~(\ref{TunnelingProbability}). In all panels, we have used the parameters $\varepsilon=5$ and $\chi_\mathrm{2D}=6.5$ {\AA}.}\label{fig:PairCorrelation}
\end{figure}

\begin{figure}[t]
\centering
\includegraphics*[width=8.5cm]{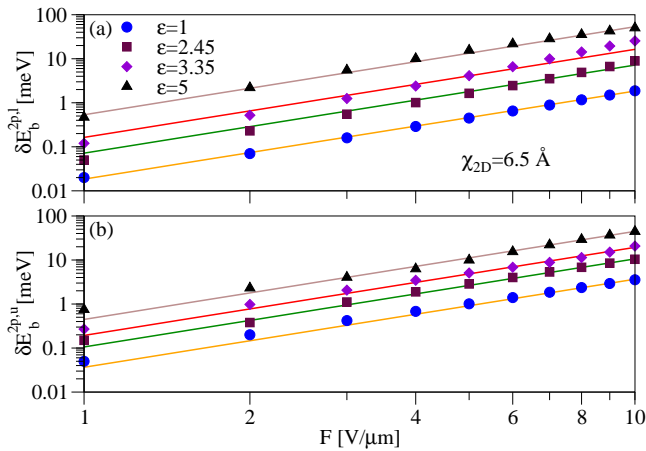}
\caption{(Color online) Field dependence of the Stark shifts of the (a) lower and (b) upper $2p$ exciton states, $\delta E^{2p,\mathrm{l}}_\mathrm{b}$ and $\delta E^{2p,\mathrm{u}}_\mathrm{b}$, for different dielectric environments. Symbols denote computed data points, whereas the solid lines denote fits to Eq.~(\ref{QuadraticStarkShift}) and Table~\ref{tab:BindingEnergies2p}.}\label{fig:pstates}
\end{figure}

Figure~\ref{fig:PairCorrelation} compares the amplitudes of the exciton wave function\footnote{In order to determine the exciton wave function, we compute the pair correlation function using the solution of the BSE~(\ref{BSE}).} $|\chi(x,y)|^2$ of the central low-energy peak for $\varepsilon=5$ and $\chi_\mathrm{2D}=6.5$ {\AA} at zero field [Fig.~\ref{fig:epsilon}~(a)] and $F=10$ V/$\mu$m [Fig.~\ref{fig:epsilon}~(b)]. Here, $x=x_\mathrm{e}-x_\mathrm{h}$ and $y=y_\mathrm{e}-y_\mathrm{h}$ denote the relative position of the electron-hole pair, and Figs.~\ref{fig:PairCorrelation}~(a)-(d) have been computed with a $240\times240$ $k$-mesh (in the full Brillouin zone). By fitting $|\chi(x,y)|^2$ to a Gaussian, we find a Bohr radius of around $11$ {\AA} for the exciton from Figs.~\ref{fig:PairCorrelation}~(a) [also shown in Fig.~\ref{fig:epsilon}~(b)], comparable to the length scales reported in the literature.\cite{Zhang2014:PRB}

If an electric field is applied, the wave function leaks out of the central region [see Figs.~\ref{fig:PairCorrelation}~(b)]. This leakage can be used to calculate the tunneling probability of an exciton into the free electron-hole continuum. For fields between $F=6$ V/$\mu$m and $F=20$ V/$\mu$m, we can fit the tunneling probability $T(F)$ to an motivated by the hydrogen atom field ionization:\cite{LandauLifshitz1981}
\begin{equation}\label{TunnelingProbability}
T(F)\propto\frac{F_0}{F}\exp\left(-F_0/F\right)
\end{equation}
with the fitting parameter $F_0=130$ V/$\mu$m. Then, $T(F)$ can in turn be related to the exciton ionization decay rate (lifetime), which is proportional (inversely proportional) to $T(F)$.\cite{Perebeinos2007:NL}

For high electric fields and/or relatively small binding energies, the exciton ionization lifetime can also be determined as follows: While an exciton and its peak at zero field corresponds to one single solution of the BSE~(\ref{BSE}), this single solution splits into several eigenstates of the BSE at finite electric field. The distribution of these split peaks, an example of which is shown in Fig.~\ref{fig:PairCorrelation}~(c), determines the intrinsic linewidth, which we find from a Gaussian fit. This procedure yields the spectrum in Fig.~\ref{fig:PairCorrelation}~(c) using a Gaussian broadening of $\Gamma_\mathrm{e}=3$ meV. The broadening $\Gamma$ from the fit in Fig.~\ref{fig:PairCorrelation}~(c) is due to the convolution of the intrinsic ionization decay rate $\Gamma_\mathrm{i}$ and an extrinsic broadening $\Gamma_\mathrm{e}$, such that $\Gamma=\sqrt{\Gamma^2_\mathrm{e}+\Gamma^2_\mathrm{i}}$. In this way, we obtain a lifetime broadening of $\Gamma_\mathrm{i}\approx2.4$ meV or $2.7\times10^{-13}$ s for $E_\mathrm{b}=143$ meV.

By adjusting the constant proportionality factor between $\Gamma_\mathrm{i}(F)$ and $T(F)$ to match the results at high fields, we can determine the ionization lifetimes also at lower fields.\cite{Perebeinos2007:NL} The calculated lifetimes are shown in Fig.~\ref{fig:PairCorrelation}~(d) with the inset showing $T(F)$ and the fit according to Eq.~(\ref{TunnelingProbability}). For lower fields, we find exciton ionization lifetimes in the ns/sub-ns range, while at higher fields the lifetimes are sub-ps. These short exciton decay lifetimes imply a rapid field-induced dissociation into free carriers that in turn can potentially contribute to photoconductivity\cite{Furchi2014:NL} and be used in photodetectors\cite{LopezSanchez2013:NN,Koppens2014:NN} or solar cells.\cite{Pospischil2014:NN}

Until now, we have only considered bright excitons that appear in the linear optical absorption as displayed in Fig.~\ref{fig:StarkAbsorption}. Labeling the excitons in analogy to the hydrogen series, the $A/B$ and $A'/B'$ excitons correspond to $1s$ and $2s$ states, respectively. In contrast, excitons with finite orbital angular momentum are dipole-forbidden and do not contribute to the linear absorption, but can be probed by two-photon absorption measurements.\cite{Poellmann2015:NM} Consistent with recent theoretical predictions,\cite{Wu2015:PRB,Berkelbach2015:PRB} we find that at zero field the $2p$ states are more strongly bound than the $2s$ state. Moreover, the two $2p$ states (per valley) are not degenerate, but split by 22 meV for $\varepsilon=1$. The upper $2p$ state, that is, the $2p$ state with lower binding energy, is in turn 56 meV below the $2s$ state for $\varepsilon=1$.

\begin{widetext}
\begin{center}
\begin{table}
\begin{center}
\begin{tabular}{|c|c||c|c|c|c|c|c|}
\hline
$\chi_\mathrm{2D}$ [\AA] & $\varepsilon$ & $E^{2p,\mathrm{l}}_\mathrm{b}$ [meV] & $\kappa_{2p,\mathrm{l}}$ [(eV$\mu$m/V)$^2$] & $\tilde{\alpha}_{2p,\mathrm{l}}$ [eV$\mu$m$^2$/V$^2$] & $E^{2p,\mathrm{u}}_\mathrm{b}$ [meV] & $\kappa_{2p,\mathrm{u}}$ [(eV$\mu$m/V)$^2$] & $\tilde{\alpha}_{2p,\mathrm{u}}$ [eV$\mu$m$^2$/V$^2$]\\
\hline\hline
6.5 & 1 & 334 & $6.17\times10^{-6}$ & $3.69\times10^{-5}$ & $312$ & $1.14\times10^{-5}$ & $7.31\times10^{-5}$\\
\hline
6.5 & 2.45 & 138 & $9.90\times10^{-6}$ & $1.43\times10^{-4}$ & $126$ & $1.33\times10^{-5}$ & $2.12\times10^{-4}$\\
\hline
6.5 & 3.35 & 97 & $1.59\times10^{-5}$ & $3.27\times10^{-4}$ & $83$ & $1.61\times10^{-5}$ & $3.88\times10^{-4}$\\
\hline
6.5 & 5 & 49 & $2.59\times10^{-5}$ & $1.07\times10^{-3}$ & $45$ & $2.00\times10^{-5}$ & $8.95\times10^{-4}$\\
\hline
\end{tabular}
\end{center}
\caption{Binding energies $E^{2p,\mathrm{l}}_\mathrm{b}$ and $E^{2p,\mathrm{u}}_\mathrm{b}$ as well as fitting parameters $\kappa_{2p,\mathrm{l}}$ and $\kappa_{2p,\mathrm{u}}$ in Eq.~(\ref{QuadraticStarkShift}) with corresponding electric polarizabilities $\tilde{\alpha}_{2p,\mathrm{l}}$ and $\tilde{\alpha}_{2p,\mathrm{u}}$ of the lower (l) and upper (u) $2p$ exciton states for the dielectric environments used in Fig.~\ref{fig:pstates}.}\label{tab:BindingEnergies2p}
\end{table}
\end{center}
\end{widetext}

Figure~\ref{fig:pstates} displays the Stark shifts $E^{2p,\mathrm{l}}_\mathrm{b}$ and $E^{2p,\mathrm{u}}_\mathrm{b}$ of the lower [Fig.~\ref{fig:pstates}~(a)] and upper [Fig.~\ref{fig:pstates}~(b)] $2p$ states for different dielectric environments. Similarly to the $s$-like $A/B$ and $A'/B'$ excitons, the Stark shift increases with $\varepsilon$, that is, with decreasing binding energy of the $2p$ states. Moreover, for both $2p$ states, we find $\delta E^{2p,\mathrm{l/u}}_\mathrm{b}$ to be of the order of tens of meV, and thus typically larger than $\delta E_\mathrm{b}$ for the $A$ and $B$ peaks, due to their smaller binding energies [see Table~\ref{tab:BindingEnergies2p}]. At small fields or large binding energies $E^{2p,\mathrm{l/u}}_\mathrm{b}$, we can fit $\delta E^{2p,\mathrm{l/u}}_\mathrm{b}$ to Eq.~(\ref{QuadraticStarkShift}), where we substitute $E_\mathrm{b}$ by $E^{2p,\mathrm{l/u}}_\mathrm{b}$ and summarize the fitting parameters in Table~\ref{tab:BindingEnergies2p}. This nonlinear behavior of $\delta E^{2p,\mathrm{l/u}}_\mathrm{b}$ is a consequence of the broken degeneracy of the $2p$ states and provides another striking difference of the excitonic series in TMDs compared to the hydrogen series.

\section{Conclusions}
We have theoretically investigated excitons and the absorption spectra of MoS$_2$ monolayers in the presence of an applied in-plane electric field using a tight-binding and Bethe-Salpeter-equation approach. Our calculations predict a quadratic Stark shift for the main exciton peaks in the linear absorption spectra, which is of the order of a few meV for fields of 10 V/$\mu$m and can exceed 30 meV for a larger electric field of 100 V/$\mu$m. Moreover, the loss of oscillator strength with the field and the scaling with the binding energy and the dielectric environment have been investigated. Our results imply that very large fields are required beyond which these excitons fully dissociate into free electron-hole pairs. Finally, we have investigated the Stark effect not only on bright excitons that appear in the linear absorption, but also on the dark $2p$ excitons. For those excitons, we predict a Stark shift of the order of tens of meV, and thus typically larger than the shift of the main absorption peaks. Remarkably, we predict the $2p$ excitons to also exhibit a nonlinear scaling, in contrast to the linear Stark effect of $p$ states in the hydrogen atom.

As the binding energies of the bright and dark excitons can be modified by placing MoS$_2$ on different substrates and thus in different dielectric environments, our results can provide theoretical guidance for a versatile substrate engineering of the electro-optical response. While we have focused on MoS$_2$, our findings suggest that such engineering should be possible in other transition metal dichalcogenide monolayers as well.

\acknowledgments
We gratefully acknowledge Rupert Huber and Yang-Fang Chen for stimulating discussions and suggestions. This work was supported by U.S. DOE, Office of Science BES, under Award DE-SC0004890 (B.S., I.\v{Z}.), and by DFG Grants No. SCHA 1899/1-1 and No. SCHA 1899/2-1 (B.S.), No. GRK 1570 (T.F., J.F.) and No. SFB 689 (M.G., J.F.).\\
\noindent{\textit{Note added.}} After the submission of our manuscript, we became aware of Refs.~\onlinecite{Haastrup2016:PRB} and~\onlinecite{Pedersen2016:PRB}, which also study the Stark shift due to an in-plane electric field.

\bibliography{BibOptics2D}




\end{document}